\documentclass[lettersize,journal]{IEEEtran}
\IEEEoverridecommandlockouts

\usepackage{amssymb}
\usepackage[cmex10]{amsmath}
\usepackage{stfloats}
\usepackage{graphicx}
\usepackage{subfigure}
\usepackage{subcaption}
\usepackage{tabularx}
\usepackage{epsfig,epsf,color,balance,cite}
\usepackage{verbatim}
\usepackage{url}
\usepackage{bm}
\usepackage{booktabs}
\usepackage{siunitx}
\usepackage{amsmath}
\allowdisplaybreaks[2]
\usepackage{ntheorem}

\usepackage{algorithm}
\usepackage{algorithmic}

\usepackage{color}
\definecolor{myc1}{rgb}{0,0,0}

\begin{document}

\title{Scene Graph-Aided Probabilistic Semantic Communication for Image Transmission}

\author{Chen Zhu,
        Siyun Liang,
        Zhouxiang Zhao,
        Jianrong Bao,~\IEEEmembership{Senior Member,~IEEE,}
        Zhaohui Yang,\\
        Zhaoyang Zhang,~\IEEEmembership{Senior Member,~IEEE,}
        and Dusit Niyato,~\IEEEmembership{Fellow,~IEEE}

\thanks{Chen Zhu is with the School of Communication Engineering, Hangzhou Dianzi University, Hangzhou 310018, China, and also with the Polytechnic Institute, Zhejiang University, Hangzhou 310015, China (e-mail: zhuc@zju.edu.cn).}
\thanks{Siyun Liang is with the Polytechnic Institute, Zhejiang University, Hangzhou 310015, China (e-mail: siyunliang@zju.edu.cn).}
\thanks{Zhouxiang Zhao, Zhaohui Yang and Zhaoyang Zhang are with the College of Information Science and Electronic Engineering, Zhejiang University, and also with Zhejiang Provincial Key Laboratory of Info. Proc., Commun. \& Netw. (IPCAN), Hangzhou 310027, China (e-mails: \{zhouxiangzhao, yang\_zhaohui, ning\_ming\}@zju.edu.cn).}
\thanks{Jianrong Bao is with the School of Communication Engineering, Hangzhou Dianzi University, Hangzhou 310018, China (e-mail: baojr@hdu.edu.cn).}
\thanks{Dusit Niyato is with the School of Computing and Data Science, Nanyang Technological University, Singapore (e-mail: dniyato@ntu.edu.sg).}
}

\maketitle

\begin{abstract}
Semantic communication emphasizes the transmission of meaning rather than raw symbols. It offers a promising solution to alleviate network congestion and improve transmission efficiency. In this paper, we propose a wireless image communication framework that employs probability graphs as shared semantic knowledge base among distributed users. High-level image semantics are represented via scene graphs, and a two-stage compression algorithm is devised to remove predictable components based on learned conditional and co-occurrence probabilities. At the transmitter, the algorithm filters redundant relations and entity pairs, while at the receiver, semantic recovery leverages the same probability graphs to reconstruct omitted information. For further research, we also put forward a multi-round semantic compression algorithm with its theoretical performance analysis. Simulation results demonstrate that our semantic-aware scheme achieves superior transmission throughput and satiable semantic alignment, validating the efficacy of leveraging high-level semantics for image communication.
\end{abstract}


\IEEEpeerreviewmaketitle

\section{Introduction}\label{sec:introduction}
\subsection{Background and Motivation}
\IEEEPARstart{O}{ver} the past decade, wireless communication has undergone significant evolution, fueled by the explosive growth of mobile internet, cloud computing, and artificial intelligence. This progression has been characterized by increased demand for data throughput, lower latency, and more reliable connectivity. The upcoming sixth generation (6G) communication networks are expected to further push these boundaries, enabling transformative applications such as extended reality (XR), high-fidelity holographic communication, unmanned aerial vehicle coordination, autonomous driving, and remote robotic surgery \cite{8869705}. These emerging use cases not only require high data rates and ultra-reliable low-latency communication (URLLC)\cite{8472907}, but also necessitate intelligent processing and comprehension of the transmitted information \cite{10550151}.

Current communication systems are predominantly anchored in classical information theory, as formulated by Shannon \cite{b1}. This framework's primary objective is the reliable and efficient transmission of symbols. Decades of theoretical innovation and technological breakthroughs have propelled modern communication systems towards the performance limits dictated by information theory. For example, in source coding, techniques like distributed compressed sensing\cite{9057744} have achieved near-optimal rate-distortion performance. Similarly, channel coding schemes such as Turbo codes\cite{397441} and low-density parity-check (LDPC)\cite{1057683} codes have approached Shannon's channel capacity limits. However, the physical layer is increasingly constrained by diminishing returns in spectral efficiency improvements and hardware resource limitations. With transmission resources finite and nearing saturation, next-generation mobile systems necessitate a departure from traditional paradigms. This entails not only addressing fundamental challenges like dynamic air interface resource scheduling\cite{4526000} but also developing intelligent resource allocation mechanisms to provide differentiated quality of service (QoS) for a massive and diverse user base\cite{6168145}. Such advancements are pivotal for transcending the performance limitations of current systems and tackling the pressing challenges in the trajectory towards 6G.

Semantic communication represents a novel architecture that integrates task relevance and user intent into the communication process \cite{shi2021semantic,11006980}. By emphasizing the effective exchange and clear conveyance of meaning, semantic communication promises to significantly enhance efficiency and user experience. This pioneering paradigm also holds the potential to resolve complex compatibility issues inherent in traditional symbol-based systems across disparate systems, protocols, and network domains\cite{gunduz2022beyond,10915662,lin2023blockchain}.

Images are a dominant information modality in daily communication, rendering them a crucial application domain for semantic communication. The rapid advancements in deep learning, particularly in natural language processing (NLP), speech recognition, and computer vision\cite{10511442}, have laid the groundwork for realizing image semantic communication. joint source-channel coding (JSCC) has emerged as a viable strategy for capturing and transmitting semantic features effectively \cite{b42}. Concurrently, various semantic representation methods for image sources have been explored, with scene graphs gaining prominence for their ability to encapsulate high-level image semantics and offer novel perspectives on image understanding \cite{b41}. Nevertheless, research in image semantic communication is nascent. As a cornerstone of next-generation communication, semantic information theory requires both rigorous mathematical formalization and practical, implementable solutions. While many contemporary approaches leverage deep learning, their underlying mechanisms and interpretability often remain opaque.

\subsection{Related Work}
The convergence of artificial intelligence and advancements in computational hardware has provided sufficient capabilities to construct sophisticated image semantic communication systems. Current research predominantly follows two main trajectories: integrated semantic and channel coding, and dedicated semantic information extraction.

\subsubsection{Joint Source-Channel Coding (JSCC)}
Classical communication theory advocates for separate source-channel coding (SSCC), where source and channel coding are designed and optimized independently. While optimal under idealized conditions (e.g., infinite code length, stationary channels), SSCC often falls short in practical scenarios characterized by finite code lengths, computational constraints, time-varying channels, and multi-user interference.
Recent advancements in AI, particularly deep learning, have fueled the development of learnable, data-driven JSCC methods
\cite{9541059}. These methods excel at extracting intrinsic source correlations and can enhance system robustness through end-to-end optimization. For image transmission, the DeepJSCC scheme, proposed by Gunduz et al. \cite{8723589}, leverages convolutional neural networks (CNNs) to directly map image pixels to channel symbols. DeepJSCC has demonstrated superior data compression and noise resilience, especially in low signal-to-noise ratio (SNR) regimes, outperforming traditional methods like JPEG and JPEG2000\cite{RABBANI20023} by avoiding the ``cliff effect" and exhibiting graceful degradation. Subsequent refinements include AD-DeepJSCC for improved SNR generalization \cite{9438648}, DeepJSCC-A for adaptive rate control \cite{9746335}, and DeepJSCC-V for enhanced transmission quality \cite{10015684}.
Despite these successes, existing DeepJSCC frameworks primarily operate on low-level image features, lacking a deeper understanding of the transmitted image content. Furthermore, their achievable compression ratios are often limited, constraining flexibility and bandwidth efficiency.

\subsubsection{Image Semantic Encoding}
Extracting meaningful semantic information from image sources is a vibrant research area. Deep learning models, with their potent feature learning and representation capabilities, have been instrumental in this domain\cite{9693341}.
CNN-based deep models\cite{7298965,7780459} are proficient at extracting low-level semantic features, forming the backbone of many computer vision tasks like image classification\cite{9084237} and object recognition\cite{7814044}. As noted, JSCC techniques also utilize CNNs for multi-granularity semantic feature extraction.
Generative adversarial networks (GANs) have proven effective for mapping image semantics to compact feature representations \cite{9879379}. Systems employing GANs typically prioritize efficient image compression, aiming for high reconstruction quality at high compression ratios. This often involves selectively omitting redundant semantic information at the transmitter and relying on the receiver to infer missing details based on the received semantics. For instance, SemanticStyleGAN \cite{9879379} extracts disentangled latent semantic features (e.g., skin tone, hairstyle) from facial images, enabling targeted reconstruction at the receiver. This approach has shown advantages over DeepJSCC in transmission efficiency and image quality \cite{10757462}.
More recent efforts focus on transforming images into textual descriptions, such as scene graphs, to achieve even higher compression ratios \cite{park2024transmitneedtaskadaptivesemantic}. Scene graphs, which represent entities and their relationships as directed graphs, are considered an ideal representation for high-level image semantics, particularly in contexts like visual storytelling \cite{park2024transmitneedtaskadaptivesemantic}.

\subsection{Contributions and Paper Structure}
This paper aims to develop a distributed image semantic communication system founded on probability graphs (PGs). By leveraging the expressive power of PGs, we seek to further compress the high-level semantic content of image sources, thereby reducing transmission overhead and enhancing overall efficiency.

The primary contributions are:
\begin{itemize}
    \item We propose a novel framework where PGs, trained on scene graph annotations from the Visual Genome dataset, serve as a shared knowledge base between the transmitter and receiver.
    \item At the transmitter, a ResNet-Transformer architecture extracts image semantics into scene graphs. A PG-based semantic compression algorithm is then applied to prune redundant relationships and entities based on their statistical prevalence in the shared knowledge base.
    \item At the receiver, the PG facilitates the recovery of elided semantic information from the compressed scene graph. A Latent Diffusion Model (LDM) subsequently reconstructs the image from the restored semantics.
    \item We introduce a distributed PG construction methodology, allowing multiple users to contribute to and benefit from a collectively enhanced shared knowledge base, improving compression efficacy, especially with limited local data.
    \item The system's feasibility and performance are validated under simplified communication assumptions, focusing on compression ratios and reconstructed image quality, demonstrating the potential of high-level semantic transmission in resource-constrained image communication.
    \item To further achieve better compression performance, we analyze the nature of a multi-round semantic compression algorithm, and perform simulation to verify the theoretical assumption of its power consumption. 
\end{itemize}

The remainder of this paper is organized as follows: Section \ref{sec:system_construction} details the architecture of the proposed image semantic communication system, including PG construction, semantic encoding, PG-based compression with the corresponding recovery algorithms and image reconstruction, and the proposed multi-round semantic compression algorithm. Section \ref{sec:experiments} presents the experimental setup, simulation results, and performance analysis of both the PG-based compression and the multi-round semantic compression. Finally, Section \ref{sec:conclusion} concludes the paper and discusses potential avenues for future research.

\section{System Construction}
\label{sec:system_construction}
\subsection{Image Semantic Information}
Image semantics can be conceptualized at multiple hierarchical levels: visual, object, and conceptual\cite{Jrgensen1995ImageAA}. The visual level pertains to low-level features like color, texture, and shape. The object level describes the attributes and states of discernible objects within the image. The conceptual level, representing the highest tier of abstraction, captures the essence of the image content in a manner akin to human understanding. For example, an image containing sand, blue sky, and water would be processed at the visual level by distinguishing these elemental features. At the object level, they would be identified as ``sand," ``blue sky," and ``water." At the conceptual level, the overarching concept "beach" might be inferred as the core semantic meaning.

While much research has focused on the visual level, particularly contextual features arising from pixel-level correlations, our system prioritizes the conceptual level, which we term high-level semantic information. Specifically, we employ scene graphs \cite{park2024transmitneedtaskadaptivesemantic} to represent these semantics. For an image $x \in \mathbb{R}^{M \times N}$, its high-level visual semantics are formalized as:

\begin{itemize}
    \item \textbf{Objects ($O$)}: Identifiable and classifiable items or entities within the image, $O(x) = \{o_i\}_{i=1}^{I}$, where $I$ is the number of identified objects, and $o_i$ is an object index or textual label (e.g., "man").
    \item \textbf{Relationships ($R$)}: Interactions and connections between objects, providing contextual information, $R(x) = \{r_{ij}\}_{1 \le i,j \le I, i \ne j}$, where $r_{ij}$ is a textual description (e.g., "wearing"). Let $|R(x)|$ denote the number of relationships.
    \item \textbf{Layouts ($L$)}: Spatial arrangements of objects, crucial for conveying image structure. This includes bounding boxes for object localization, $L(x) = \{l_i\}_{i=1}^{I}$, where $l_i = \{(m,n) \in \mathbb{R}^{M \times N} | (m,n) \text{ belongs to } o_i\}$.
    \item \textbf{Scene Graph ($G$)}: A structured representation of objects and their relationships, $G(x)=(V,E)$, where nodes $V=\{v_i\}$ correspond to objects ($v_i=o_i$), and edges $E=\{e_{ij}\}$ represent relationships ($e_{ij}=r_{ij}$). A scene graph can be decomposed into a set of relational triplets: $G(x)=\{g_k=(o_i, r_{ij}, o_j) | o_i, o_j \in O, r_{ij} \in R\}$. Each triplet can be viewed as a concise statement (e.g., "man0 wearing hat0").
\end{itemize}

\subsection{System Framework}
\begin{figure*}[t]
\centering
\includegraphics[width=\linewidth]{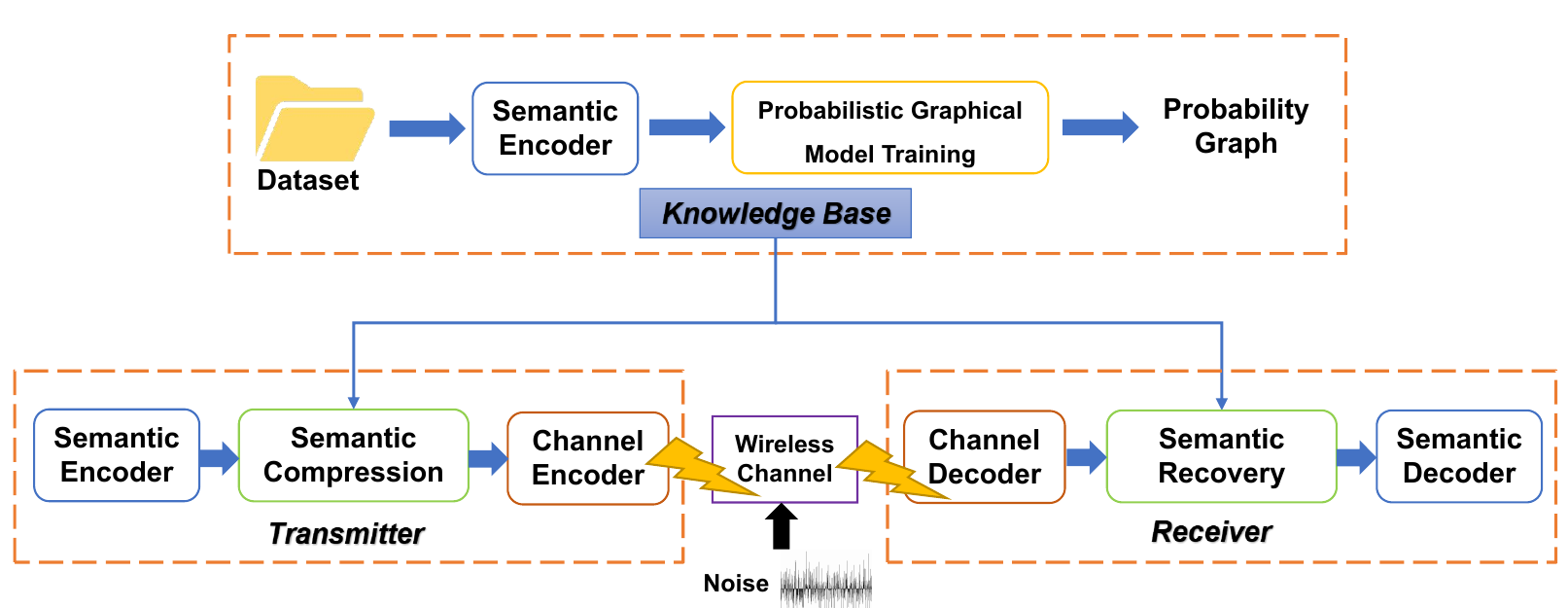}
\caption{Overall architecture of the proposed probabilistic graphical model-based image semantic communication system. The system comprises a shared semantic knowledge base (PG) and interconnected transmitter and receiver modules.}
\label{fig:system_architecture}
\end{figure*}

The proposed communication architecture is depicted in Fig.~\ref{fig:system_architecture}.

The system comprises three main components: a shared semantic knowledge base, the semantic transmitter and semantic receiver. 
\begin{itemize}
    \item \textbf{Shared semantic knowledge base:}The knowledge base is realized as PGs \cite{10915662}, trained offline using semantic encodings (scene graphs) extracted from a large annotated image dataset (e.g., Visual Genome\cite{10.1007/s11263-016-0981-7}). In practice, this knowledge base can be accumulated by the transmit context of given communication endpoints. By leveraging the known information in shared semantic knowledge base, the transmit data size can be significantly reduced.
    \item \textbf{Semantic transmitter:}At the transmitter, an input image undergoes semantic extraction to produce a scene graph. This scene graph is then compressed by a PG-based semantic compressor, which leverages the shared knowledge base to eliminate redundant information. The compressed semantic representation is subsequently encoded by a channel encoder and transmitted over a noisy communication channel.
    \item \textbf{Semantic Receiver:}At the receiver, the received signal is processed by a channel decoder. A PG-based semantic recovery module then utilizes the shared PG to infer and restore the complete scene graph from the compressed representation. Finally, a semantic decoder (image generator) reconstructs the visual image based on the recovered semantic information. 
\end{itemize}  

\subsection{Semantic Extraction}
We adapt the semantic extractor architecture from \cite{park2024transmitneedtaskadaptivesemantic}. An input image $x \in \mathbb{R}^{H \times W \times 3}$ (e.g., $512 \times 512 \times 3$) is first processed by a feature extractor. In our implementation, a ResNet-50 encoder, pre-trained on ImageNet, transforms the image $x$ into a high-dimensional feature map $F(x) \in \mathbb{R}^{C \times X \times Y}$ (e.g., $C=2048, X=Y=16$ for a $512 \times 512$ input). A $1 \times 1$ convolutional layer then reduces the channel dimension to $d=256$, yielding $F_0(x) \in \mathbb{R}^{d \times X \times Y}$.

Subsequently, a transformer-based \cite{10105507} encoder-decoder architecture extracts objects and their layouts. This typically involves a stack of encoder layers that process $F_0(x)$ to produce context-aware features $F_c(x)$, and decoder layers that utilize learnable object queries to attend to $F_c(x)$ and predict a list of objects $O(x)$ along with their bounding box layouts $L(x)$. For relation and scene graph generation, a dual-branch transformer architecture is employed. This module,  trained on Visual Genome dataset, processes pairs of detected objects $(o_i, o_j)$ to predict the relationship $r_{ij}$ between them. By combining the detected objects $O(x)$ and their inferred relationships $R(x)$, the scene graph $G(x)$ is constructed as a set of triplets $\{g_k = (o_i, r_{ij}, o_j)\}$.

\subsection{Probability Graph Construction}
PGs integrate graph theory and probability theory to model and infer dependencies among random variables. They provide a visual and mathematical framework for representing complex joint probability distributions. In our system, a PG serves as a shared semantic knowledge base.

Given that scene graphs are essentially knowledge graphs composed of (head entity, relation, tail entity) triplets, we adapt methodologies for PG construction from textual knowledge graphs \cite{ZHAO2024107055}. For a collection of images sharing a common theme or context, their corresponding scene graph annotations are statistically aggregated.
Let $P_{\text{image}} = \{P_1, \ldots, P_N\}$ be a set of $N$ training images, and $T_{\text{annotation}} = \{T_1, \ldots, T_N\}$ be their corresponding scene graph annotations. Each $T_n \in T_{\text{annotation}}, n\in \{1,2,\ldots,N\}$ can be represented as a set of triplets $G_n = \{\kappa_{nm} = (h_{nm}, r_{nm}, t_{nm})\}$, where $h_{nm}$ denotes the head entity, $ t_{nm}$ denotes the tail entity, and $r_{nm}$ denotes the relation between them.
By analyzing the co-occurrence statistics of entities and relations across all $G_n$, we construct a global PG. This PG captures:
\begin{enumerate}
    \item \textbf{Relation Probabilities}: For each unique head-tail entity pair $(h, t)$ observed in the training data, we compute the probability distribution over all possible relations $r$ that connect them. A quadruple $\rho_r = (h, [(r_1, N_{r1}), \ldots, (r_I, N_{rI})], t)$ stores the counts $N_{ri}$ for each relation $r_i$ between $h$ and $t$. The conditional probability $p(r_i | h, t)$ is then estimated as:
    \begin{equation}
    p(r_i | h, t) = \frac{\text{Count}(h, r_i, t)}{\sum_{j=1}^{I} \text{Count}(h, r_j, t)} \label{eq:rel_prob}
    \end{equation}
    This results in a relation probability graph $G_{rel\_prob}$.
    \item \textbf{Tail Entity Co-occurrence Probabilities}: For each unique head entity $h$, we compute the probability distribution over all tail entities $t$ that co-occur with it, irrespective of the specific relation. A structure $\rho_{h} = (h, [(t_1, N_{h,t1}), \ldots, (t_T, N_{h,tT})])$ stores the co-occurrence counts. The conditional probability $p(t_j | h)$ is estimated as:
    \begin{equation}
    p(t_j | h) = \frac{\text{Count}(h, t_j)}{\sum_{k=1}^{T} \text{Count}(h, t_k)} \label{eq:tail_entity_prob}
    \end{equation}
    This yields an entity co-occurrence probability graph $G_{co}$.
\end{enumerate}
These probabilities form the core of our shared PG, enabling informed decisions during semantic compression and recovery.

\begin{figure}[t]
\centering
\includegraphics[width=0.9\linewidth]{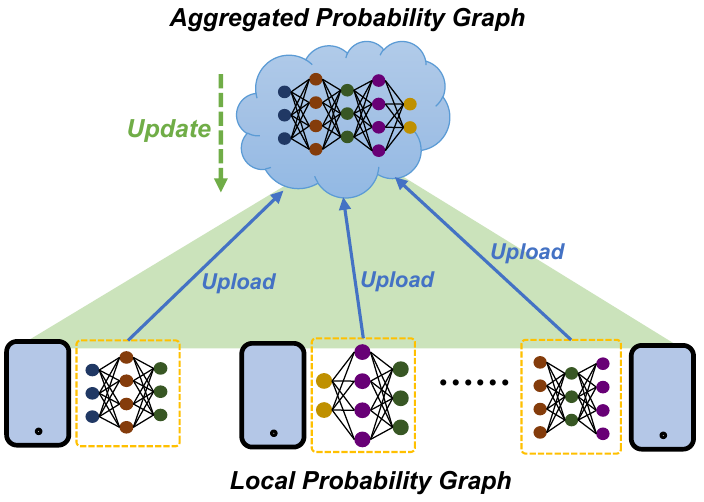}
\caption{Schematic of distributed PG construction involving multiple users. Local PGs are trained by users and then aggregated into a shared PG, which is subsequently updated and distributed.}
\label{fig:two_user_pgm}
\end{figure}

We also envision a distributed PG construction architecture, as illustrated in Fig.~\ref{fig:two_user_pgm}. Individual users can train local PGs ($G_1, G_{co1}$ for User 1; $G_2, G_{co2}$ for User 2) based on their specific data and usage patterns. These local PGs can then be periodically uploaded to a central server, which aggregates them to create an enriched shared PG ($G, G_{co}$). This shared PG is subsequently disseminated back to the users. This federated learning-inspired approach allows the system to adapt to diverse user contexts and reduce the computational burden on the central server, while leveraging collective knowledge.

\subsection{Semantic Compression and Recovery}
\subsubsection{Semantic Compression Algorithm}
Leveraging the shared PG, the extracted scene graph $G(x)$ undergoes a multi-round compression process at the transmitter to reduce data payload.

\textbf{Stage 1: Relation pruning.} This stage focuses on pruning redundant relations within triplets $(h, r, t)$. Based on the relation probability graph $G_{rel\_prob}$, if the conditional probability $p(r | h, t)$ for the most frequent relation between a given head $h$ and tail $t$ exceeds a predefined threshold $\theta_r$ (e.g., 0.5), that specific relation $r$ is considered highly predictable and is omitted from the triplet, leaving only the pair $(h, t)$. Less probable relations are retained. 

\textbf{Stage 2: Tail Entity Filtering.} This stage operates on the pairs $(h, t)$ resulting from Stage 1 (where the relation was omitted) or on original pairs if relations were not considered. It aims to prune predictable tail entities. Using the entity co-occurrence PG $G_{co}$, if the tail entity $t$ in a pair $(h,t)$ is the most probable entity to co-occur with head $h$ (i.e., $t = \arg\max_{t'} p(t' | h)$) and $p(t|h)$ exceeds a threshold $\theta_t$, then $t$ is omitted, leaving only the head entity $h$. This is described in Algorithm \ref{alg:compression}.

\begin{algorithm}[htbp]
\caption{Two-Stage Semantic Compression}
\label{alg:compression}
\begin{algorithmic}[1]
\REQUIRE Scene graph $\mathcal{G} = \{(h, r, t)\}$, global PG $\mathcal{M}_G$
\ENSURE Compressed tokens $\mathbf{s}_c$
\STATE $\mathcal{G}_{\text{filter1}} \gets \emptyset$
\FOR{$(h, r, t) \in \mathcal{G}$}
    \IF{$P_{\mathcal{M}_G}(r | h, t) < 0.5$} \label{line:relation_prune}
        \STATE $\mathcal{G}_{\text{filter1}} \gets \mathcal{G}_{\text{filter1}} \cup \{(h, r, t)\}$
    \ELSE
        \STATE $\mathcal{G}_{\text{filter1}} \gets \mathcal{G}_{\text{filter1}} \cup \{(h, t)\}$ \label{line:entity_pair}
    \ENDIF
\ENDFOR
\STATE $\mathcal{G}_{\text{filter2}} \gets \emptyset$
\FOR{$(h, t) \in \mathcal{G}_{\text{filter1}}$}
    \IF{$t = \arg\max_{t'} P_{\mathcal{M}_G}(t' | h)$} \label{line:entity_prune}
        \STATE $\mathcal{G}_{\text{filter2}} \gets \mathcal{G}_{\text{filter2}} \cup \{h\}$
    \ELSE
        \STATE $\mathcal{G}_{\text{filter2}} \gets \mathcal{G}_{\text{filter2}} \cup \{(h, t)\}$
    \ENDIF
\ENDFOR
\STATE Encode $\mathcal{G}_{\text{filter2}}$ into $\mathbf{s}_c$ via Huffman coding
\end{algorithmic}
\end{algorithm}

\begin{figure*}[t]
\centering
\includegraphics[width=\linewidth]{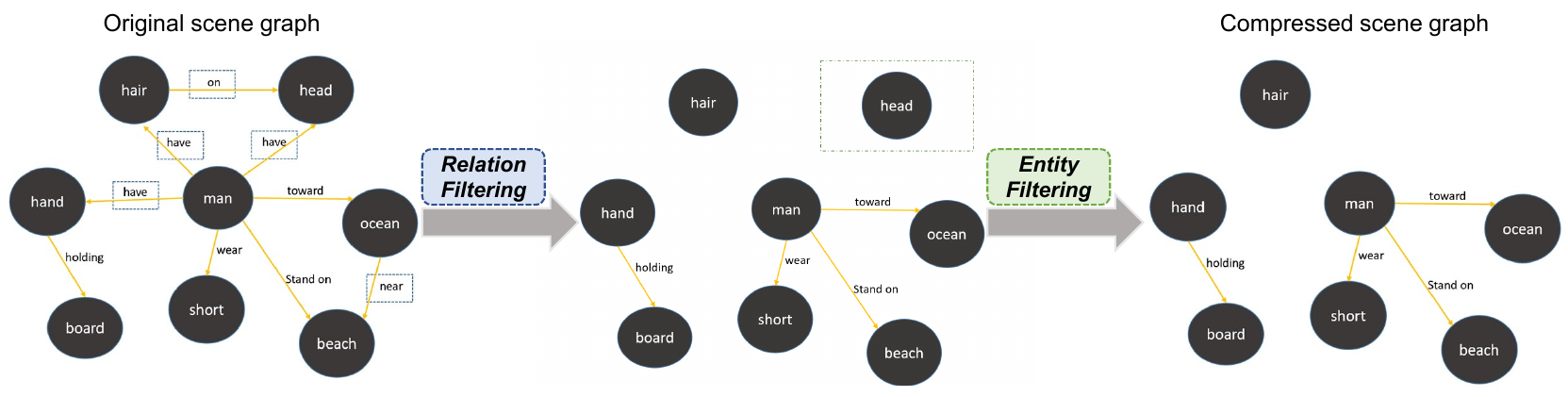}
\caption{Illustration of scene graph filtering.}
\label{fig:filter1}
\end{figure*}

Fig.~\ref{fig:filter1} illustrates an example of these two filtering processes.


\subsubsection{Semantic Recovery Algorithm}
At the receiver, the compressed semantic information $FilteredSG_2$ is restored by reversing the compression steps, leveraging the shared PGs $G_{rel\_prob}$ and $G_{co}$. This aims to reconstruct the original scene graph as accurately as possible.

\textbf{Step 1: Tail Entity Recovery.} For each lone head entity $h$ in $FilteredSG_2$, the most probable co-occurring tail entity $t_{pred} = \arg\max_{t'} p(t' | h)$ is inferred from $G_{co}$ and appended to form the pair $(h, t_{pred})$.

\textbf{Step 2: Relation Recovery.} For each pair $(h, t)$ (either originally transmitted or recovered in Step 1), the most probable relation $r_{pred} = \arg\max_{r'} p(r' | h, t)$ is inferred from $G_{rel\_prob}$ to form the triplet $(h, r_{pred}, t)$. Triplets that were transmitted with explicit relations bypass this step.
The complete recovery process is outlined in Algorithm \ref{alg:recovery}.

\begin{algorithm}[htbp]
\caption{Semantic Recovery Algorithm}
\label{alg:recovery}
\begin{algorithmic}[1]
\REQUIRE Compressed tokens $\mathbf{s}_c$, global PG $\mathcal{M}_G$
\ENSURE Reconstructed scene graph $\hat{\mathcal{G}}$
\STATE Decode $\mathbf{s}_c$ to $\mathcal{G}_{\text{filter2}}$
\STATE $\hat{\mathcal{G}}_{\text{filter1}} \gets \emptyset$
\FOR{$s \in \mathcal{G}_{\text{filter2}}$}
    \IF{$s$ is single entity $h$}
        \STATE $t \gets \arg\max_{t'} P_{\mathcal{M}_G}(t' | h)$
        \STATE $\hat{\mathcal{G}}_{\text{filter1}} \gets \hat{\mathcal{G}}_{\text{filter1}} \cup \{(h, t)\}$
    \ELSE
        \STATE $\hat{\mathcal{G}}_{\text{filter1}} \gets \hat{\mathcal{G}}_{\text{filter1}} \cup \{s\}$
    \ENDIF
\ENDFOR
\STATE $\hat{\mathcal{G}} \gets \emptyset$
\FOR{$(h, t) \in \hat{\mathcal{G}}_{\text{filter1}}$}
    \STATE $r \gets \arg\max_{r} P_{\mathcal{M}_G}(r | h, t)$
    \STATE $\hat{\mathcal{G}} \gets \hat{\mathcal{G}} \cup \{(h, r, t)\}$
\ENDFOR
\end{algorithmic}
\end{algorithm}
\subsection{Multi-round Semantic Compression}
\subsubsection{Algorithm Description}
This semantic compression algorithm operates through iterative rounds to progressively reduce data redundancy by leveraging probabilistic dependencies in a shared knowledge graph $\mathcal{G}$. In the first round, the base station extracts semantic triples $\mathcal{K} = \{k_1, \dots, k_N\}$ from raw data, where each triple $k_n = \langle h_n, t_n, r_n \rangle$ consists of a head entity $h_n$, a tail entity $t_n$, and their relationship $r_n$. During traversal, triples are categorized into three subsets: $\mathcal{K}_{\text{new}}$ (retained for transmission if $\mathcal{G}$ lacks their entity pairs or relationships), $\mathcal{K}_{\text{omit}}$ (omitted if $\mathcal{G}$ confirms their relationships as highest-probability entries), and $\mathcal{K}_{\text{mid}}$ (intermediate triples requiring further analysis).

Subsequent rounds refine $\mathcal{K}_{\text{mid}}$ using historical omissions. In the second round, a conditional probability matrix $\mathbf{M}^{(m)}$ is constructed for each intermediate triple $k_m$, where entries $\mathbf{M}^{(m)}_{ij}$ quantify the likelihood of relationship $r_j$ given an omitted triple $k_i$ from $\mathcal{K}_{\text{omit}}$ according to \eqref{eq:tail_entity_prob}. If $k_m$'s relationship corresponds to the maximum probability in $\mathbf{M}^{(m)}$, it is replaced with an index reference to $k_i$ and moved to $\mathcal{K}_{\text{omit}}$. For higher rounds ($n \geq 3$), a multi-dimensional probability tensor $\mathbf{T}^{(n)}$ is built using $n-1$ previously omitted triples, enabling the evaluation of complex conditional dependencies. Relationships are omitted if the tensor-derived probability of the corresponding relation is maximized with certain omitted triples, and the omitted relations will be replaced by the index of omitted triples used to construct the probability tensor. The algorithm terminates when no new omissions occur, ensuring a balance between compression efficiency and computational tractability despite the exponentially growing complexity of tensor operations. The process of the whole algorithm is given in Algorithm.~\ref{alg:semantic_compression}.
\begin{algorithm}
\caption{Multi-round Semantic Compression}\label{alg:semantic_compression}
\begin{algorithmic}[1]
\REQUIRE Semantic triples $\mathcal{K} = \{k_1, \dots, k_N\}$, Shared probabilistic graph $\mathcal{G}$
\ENSURE Compressed triples $\mathcal{K}_{\text{new}}$, $\mathcal{K}_{\text{omit}}$

\STATE \textbf{First Round Compression:}
\FOR{each $k_n = \langle h_n, t_n, r_n \rangle \in \mathcal{K}$}
    \IF{$(h_n, t_n) \in \mathcal{G}$ \AND $\arg\max_r P(r|h_n,t_n) = r_n$}
        \STATE Add $k_n$ to $\mathcal{K}_{\text{omit}}$
    \ELSIF{$(h_n, t_n) \notin \mathcal{G}$ \OR $r_n \notin \mathcal{G}(h_n,t_n)$}
        \STATE Add $k_n$ to $\mathcal{K}_{\text{new}}$
    \ELSE
        \STATE Add $k_n$ to $\mathcal{K}_{\text{mid}}$
    \ENDIF
\ENDFOR

\STATE \textbf{Second Round Compression:}
\FOR{each $k_m = \langle h_m, t_m, r_m \rangle \in \mathcal{K}_{\text{mid}}$}
    \STATE Query $\mathcal{G}$ to get $\langle h_m, t_m, R_m, P_m \rangle$
    \STATE Build conditional probability matrix $\mathbf{M}^{(m)}$.
    \IF{$\exists k_i \in \mathcal{K}_{\text{omit}}$ s.t. $\mathbf{M}^{(m)}_{ij}$ is maximal for $r_j = r_m$}
        \STATE Replace $r_m$ with index of $k_i$, add $k_m$ to $\mathcal{K}_{\text{omit}}$
    \ELSE
        \STATE Retain $k_m$ in $\mathcal{K}_{\text{mid}}$
    \ENDIF
\ENDFOR

\STATE \textbf{Higher-Round Compression:}
\WHILE{$|\mathcal{K}_{\text{omit}}^{(t)}| > |\mathcal{K}_{\text{omit}}^{(t-1)}|$}
    \STATE Construct tensor $\mathbf{T}^{(t)}$ using $\{\mathcal{K}_{\text{omit}}^{(1)}, \dots, \mathcal{K}_{\text{omit}}^{(t-1)}\}$
    \FOR{each $k_m^{(t)}=\langle h_m^{(t)}, t_m^{(t)}, r_m^{(t)} \rangle \in \mathcal{K}_{\text{mid}}^{(t-1)}$}
        \IF{$\mathbf{T}^{(t)}_{:,\dots,:,r_m^{(t)}})$ is maximized with $r_m^{(t)}$}
            \STATE Replace the relation of $k^{(t)}$ with indices of $\mathbf{T}^{(t)}$, add to $\mathcal{K}_{\text{omit}}$
        \ENDIF
    \ENDFOR
\ENDWHILE

\RETURN $\mathcal{K}_{\text{new}}, \mathcal{K}_{\text{omit}}$
\end{algorithmic}
\end{algorithm}
\subsubsection{Theoretical Analysis}
To show the impact of semantic compression, semantic compression ratio $\rho$ is adopted. It is defined as
\begin{equation}
    \rho=\frac{\rm{size}(\mathcal{Y})}{\rm{size}(\mathcal{X})},
\end{equation}
where $\mathcal{X}$ and $\mathcal{Y}$ denote the data before and after compression, respectively.

\begin{figure}
    \centering
    \includegraphics[width=\linewidth]{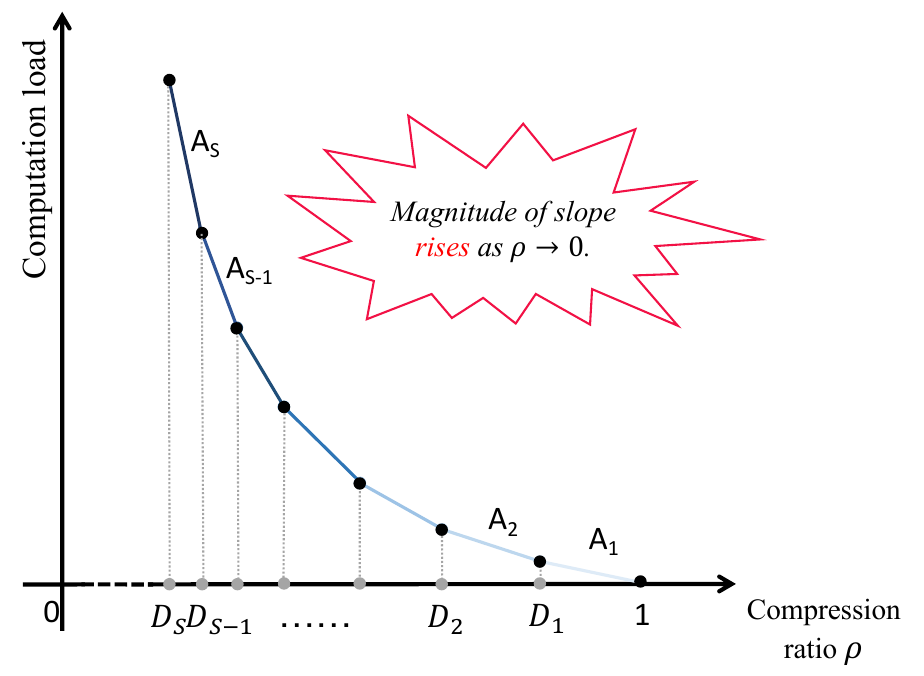}
    \caption{Illustration of computation load function.}
    \label{fig:cl}
\end{figure}

According to \cite{e26050394}, the computation load of multi-round semantic compression can be modeled as
\begin{equation}\label{eq:gn}
    c\left(\rho\right)=\left\{\begin{array}{l}
        A_1\rho +B_1, D_1\leq \rho \leq 1, \\
        A_2\rho +B_2, D_2\leq \rho < D_1, \\
        \vdots \\
        A_S\rho +B_S, D_S\leq \rho < D_{S-1},
    \end{array}\right.
    \vspace{-1mm}
\end{equation}
where $s \in \{1,2,\ldots,S\}$ is the index of each segment of the computation load function, and $B_s$ represents the intercept and $A_s$ denotes the slope of each segment, which  subjects to $0>A_1>A_2>\cdots>A_S$. $D_s$ serves as the partition of the computation load function. The nature of \eqref{eq:gn} is illustrated in Fig.~\ref{fig:cl}.

\eqref{eq:gn} allows us to analyze the power consumption of the proposed semantic compression quantitatively. To further examine the nature of \eqref{eq:gn}, a simulation using sensor-collected dataset is proposed in Section.III-D.

\subsection{Image Reconstruction}
Generative models are pivotal for transforming the recovered semantic representation (scene graph and layouts) back into a visually coherent image. We employ a Latent Diffusion Model (LDM)\cite{9878449} for this task due to its efficiency and high-quality generation capabilities. LDMs operate in a lower-dimensional latent space learned by a pre-trained Variational Autoencoder (VAE), rather than directly in pixel space. This allows the model to focus on essential semantic attributes, leading to faster and more resource-efficient generation.
The LDM is conditioned on the recovered scene graph $G_{recovered}$ and the object layouts $L(x)$. The scene graph is typically encoded into an embedding space using a transformer encoder. These conditional embeddings, along with layout information, guide the reverse diffusion process within the LDM's U-Net architecture, typically via cross-attention mechanisms. The LDM's objective function is:
\begin{equation}
\mathcal{L}_{\text{LDM}} := \mathbb{E}_{z_0, \mathbf{c}, \epsilon \sim \mathcal{N}(0,I), t} [\|\epsilon - \epsilon_{\theta}(z_t, t, \tau_{\theta}(\mathbf{c}))\|^2_2] \label{eq:ldm_loss_revised}
\end{equation}
where $z_0$ is the VAE-encoded latent representation of the original image, $z_t$ is the noisy latent at timestep $t$, $\epsilon$ is the noise, $\epsilon_{\theta}$ is the noise prediction network, $\mathbf{c}$ represents the conditioning information (scene graph and layout embeddings), and $\tau_{\theta}(\mathbf{c})$ is the domain-specific conditioning encoder. This ensures that the generated image aligns with the provided semantic and spatial constraints.

\section{Experimental Evaluation}
\label{sec:experiments}
\subsection{Experimental Setup}
\textbf{Platform}: Experiments were conducted on Google Colab Pro, utilizing NVIDIA A100 GPUs. Key software versions include Python 3.11.0, Transformers 4.46.3, OpenCV-Python 4.10.0.84, and Diffusers 0.31.0.

\textbf{Dataset}: The Visual Genome dataset was employed. Its extensive collection of images with detailed annotations of objects, attributes, and relationships makes it highly suitable for training and evaluating semantic understanding and generation models.

\textbf{Model Architecture}:
\begin{itemize}
    \item \textit{Semantic Knowledge Base}: PGs ($G_{rel\_prob}$, $G_{co}$) trained on scene graphs from the Visual Genome training split.
    \item \textit{Semantic Transmitter}: Comprises a semantic extractor (ResNet-50 + dual-branch Transformer for scene graph generation) and the PG-based semantic compressor (Algorithms \ref{alg:compression} with $\theta_r = 0.5, \theta_t = 0.5$).
    \item \textit{Channel}: A simplified Additive White Gaussian Noise (AWGN) channel was assumed for throughput evaluations.
    \item \textit{Semantic Receiver}: Consists of the PG-based semantic recovery module, as presented in Algorithm \ref{alg:recovery}, and a pre-trained LDM for image reconstruction, conditioned on recovered scene graphs and original layouts.
\end{itemize}
\textbf{Evaluation Metrics}:
\begin{itemize}
    \item \textit{Compression Ratio}: Measured as the ratio of the size of the compressed scene graph to the original scene graph.
    \item \textit{Throughput}: Number of images whose semantic information can be transmitted per second under varying SNR conditions.
    \item \textit{Image Quality}:
        \begin{itemize}
            \item \textbf{CLIP (Contrastive Language-Image Pre-Training) Similarity} \cite{CLIP}: CLIP is a multimodal model that jointly trains image and text encoders through contrastive learning, enabling the mapping of both images and text into a unified semantic space. The core concept of CLIP involves learning the associations between visual content and natural language descriptions using massive-scale image-text pairs. For instance, given an image depicting "a dog running on the grass", CLIP encodes both the image and its corresponding textual description into high-dimensional vectors while preserving their proximity within the vector space. We leverage CLIP to compute the semantic similarity between two images. Specifically, the original image and the generated image are fed into CLIP's image encoder, yielding two high-dimensional feature vectors $I_o$ and $I_g$. The cosine similarity between these vectors is then calculated, where a value closer to 1 indicates higher semantic consistency. This metric is formulated as
            \begin{equation}
                \text{CLIP}(I_o, I_g) = \frac{I_o \cdot I_g}{\|I_o\| \|I_g\|}.
            \end{equation}

            \item \textbf{LPIPS (Learned Perceptual Image Patch Similarity)} \cite{LPIPS}: LPIPS is a deep learning-based image similarity metric that employs pre-trained convolutional neural networks (e.g., VGG16, AlexNet) to extract multi-level features from images. After normalizing these features, it computes layer-wise weighted differences and aggregates them into a perceptual similarity score. Its core innovation lies in simulating human visual perception by utilizing deep network features to capture semantic and structural differences, making it more aligned with subjective evaluations than traditional metrics (e.g., PSNR, SSIM). Consequently, LPIPS is widely adopted for quality assessment in image generation, super-resolution, and related tasks.  The similarity computation involves calculating cosine distances between corresponding channels of network outputs at each layer, followed by spatial and hierarchical averaging. Specifically, LPIPS processes two feature sets through the following pipeline: For base networks \(F\), it computes deep embeddings, normalizes activations along channel dimensions, scales each channel by vector \(c\), computes the \(L_2\) distance, and finally averages across spatial dimensions and all layers. The distance \(d_0\) between image \(x\) and reference \(x_0\) is formalized in:  

\begin{equation}
d_0 = \frac{1}{H_l W_l} \sum_{h,w} \left\| c_l \odot \left( \hat{F}_l^{(x)} - \hat{F}_l^{(x_0)} \right) \right\|_2^2, 
\end{equation}
where \(\hat{F}_l\) denotes normalized features at layer \(l\), \(c_l\) represents channel-wise scaling weights, and \(H_l\), \(W_l\) are spatial dimensions.
        \end{itemize}
\end{itemize}

\subsection{Simulation Results and Analysis}
\subsubsection{Scene Graph Generation and Reconstruction for Low-semantic-density Image}
To verify the effectiveness of the communication scheme, we initially focus on images containing only simple geometric shapes. Given that predefined shapes (\texttt{triangle}, \texttt{circle}, \texttt{rectangle}) contain limited semantic information, we exclusively consider their relative spatial relationships (\texttt{left}, \texttt{right}, \texttt{above}, \texttt{below}). After determining primary spatial orientations, we generate $25$ images with the aforementioned feature, with $20$ images used to construct a probabilistic graphical model containing only prior relational distributions. In simulation, we extract the scene graphs from the rest $5$ images and perform relation filtering according to the shared probability graph. The simulation encompasses the following stages: probabilistic graphical model construction, semantic extraction and compression, semantic recovery, and image reconstruction.

\begin{figure}
    \centering
    \includegraphics[width=0.75\linewidth]{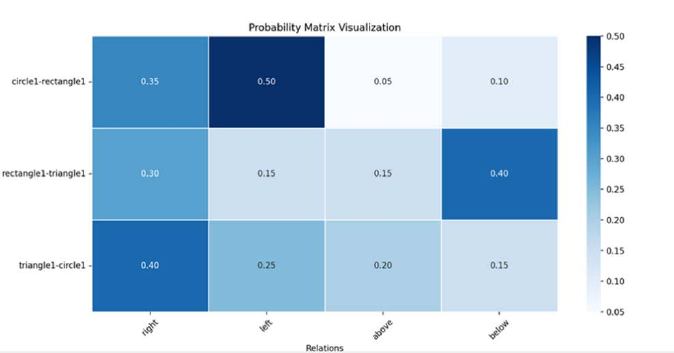}
    \caption{Probability matrix for low-semantic-density image generation.}
    \label{fig:prob_matrix}
\end{figure}

Since relationships between objects in simple-shape images are randomly generated, we focus exclusively on relationship compression. The joint probability distribution of all relationships is computed, with the constructed probabilistic graphical model characterized by the norm of the probability matrix, as illustrated in Fig.~\ref{fig:prob_matrix}.

\begin{figure}[htbp] 
		\centering  
		\vspace{-0.35cm} 
		\subfigtopskip=2pt 
		\subfigbottomskip=2pt 
		\subfigcapskip=-5pt 
		\subfigure[Layout for low-semantic-density image.]
		{
			\label{fig:layout_simp}
			\includegraphics[width=0.32\linewidth]{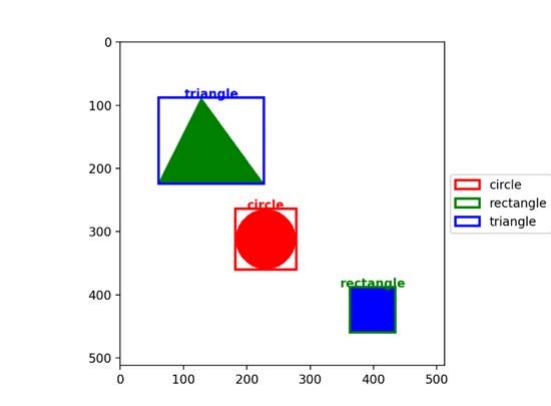}
		}
		\quad
		\subfigure[Extracted scene graph for low-semantic-density image.]
		{
			\label{fig:sg_full}
			\includegraphics[width=0.32\linewidth]{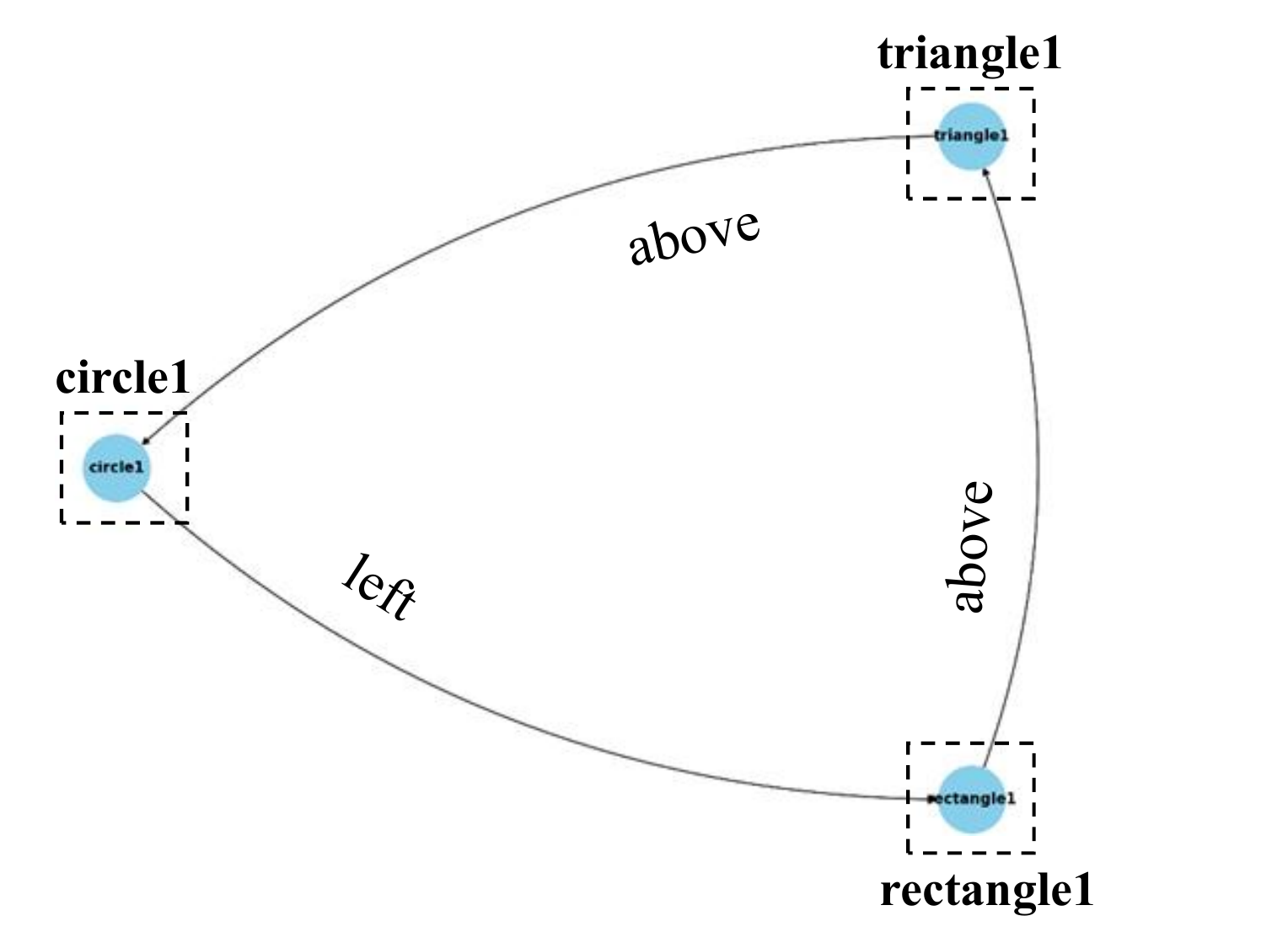}
		}
		\subfigure[Filtered scene graph for low-semantic-density image.]
		{
			\label{fig:sg_filt}
			\includegraphics[width=0.32\linewidth]{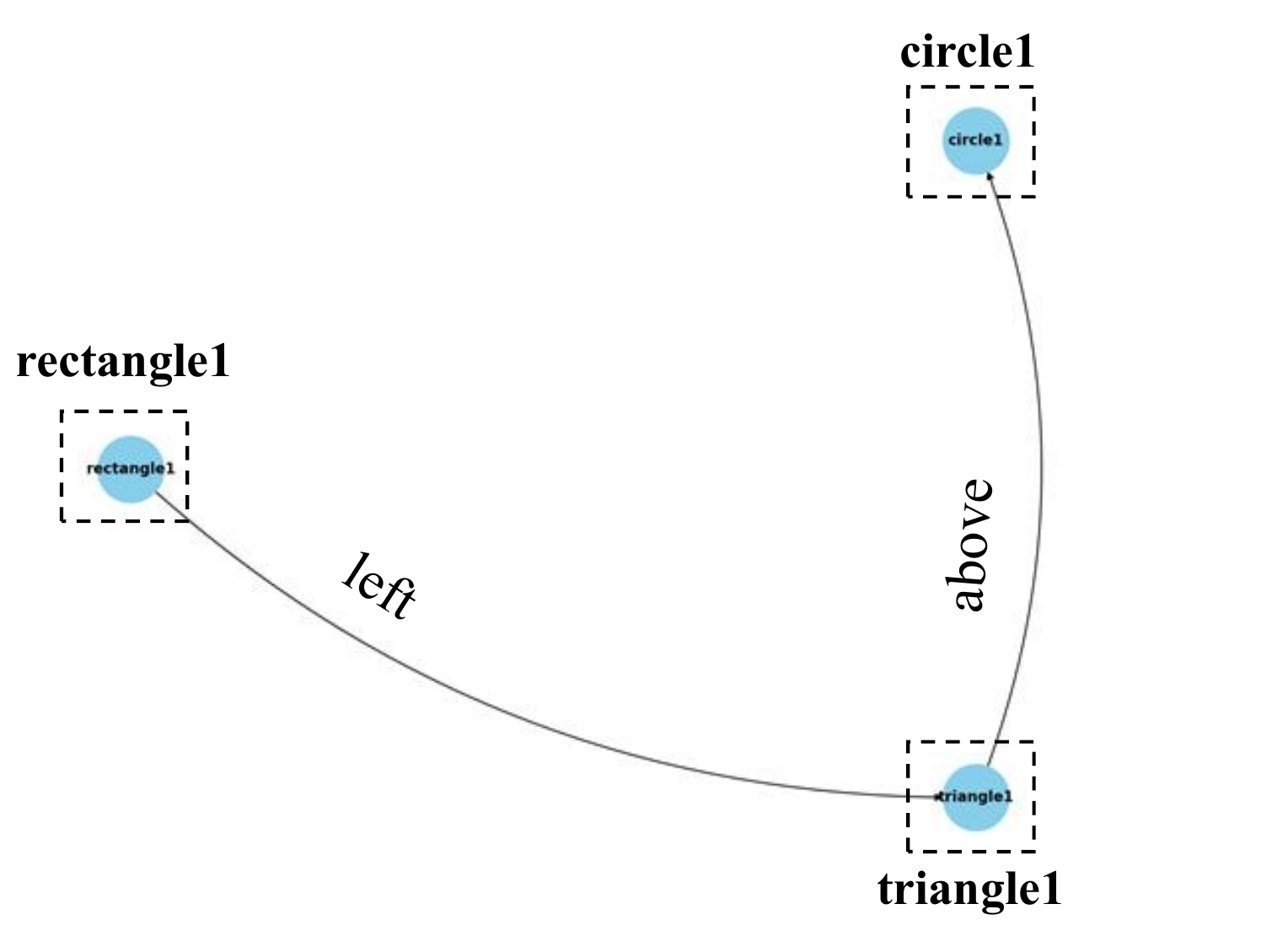}
		}
		\quad
		\subfigure[reconstructed image generated scene graph.]
		{
			\label{fig:sg_gen}
			\includegraphics[width=0.32\linewidth]{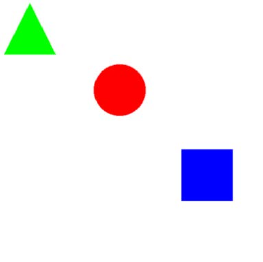}
		}
		\caption{Experiment results for scene graph generation and reconstruction for low-semantic-density image}
		\label{level}
	\end{figure}

An example of the layout of the low-semantic density image is given in Fig.~\ref{fig:layout_simp}, and the corresponding extracted scene graph is demonstrated in Fig.~\ref{fig:sg_full}.  After relation filtering, as shown in Fig.~\ref{fig:sg_filt}, the relation that has the largest conditional probability with given head, tail entity, is omitted (\texttt{above} between \texttt{triangle} and \texttt{circle}). The full scene graph can be easily re-constructed leveraging the shared probability graph (in this case, it can be derived from the probability matrix shown in Fig.~\ref{fig:prob_matrix}). The reconstructed image generated scene graph is shown in Fig.~\ref{fig:sg_gen}. After testing the reconstruction process on five test images, the calculated structural similarity index measure (SSIM) of the reconstructed images achieved an average value of 0.857. This demonstrates that in cases of simple image composition, the receiver can easily reconstruct satisfactory results with extracted semantic information only. This simulation outlines a fundamental workflow: When the semantic extractor sufficiently comprehends the transmitted image content, visually acceptable results can be achieved with minimal data transmission. Under constrained channel bandwidth or limited transmission resources, our approach involves:
\begin{itemize}
    \item Semantic extraction and probabilistic graph-based compression of visual semantic information at the source.
    \item Lost semantic information recovery through probabilistic graphical models at the receiver. This dual-process effectively fulfills transmission requirements while maintaining perceptual quality.
\end{itemize}

\subsubsection{Simulation of Transmission Workflow}
\begin{figure}[t]
\centering
\includegraphics[width=0.8\linewidth]{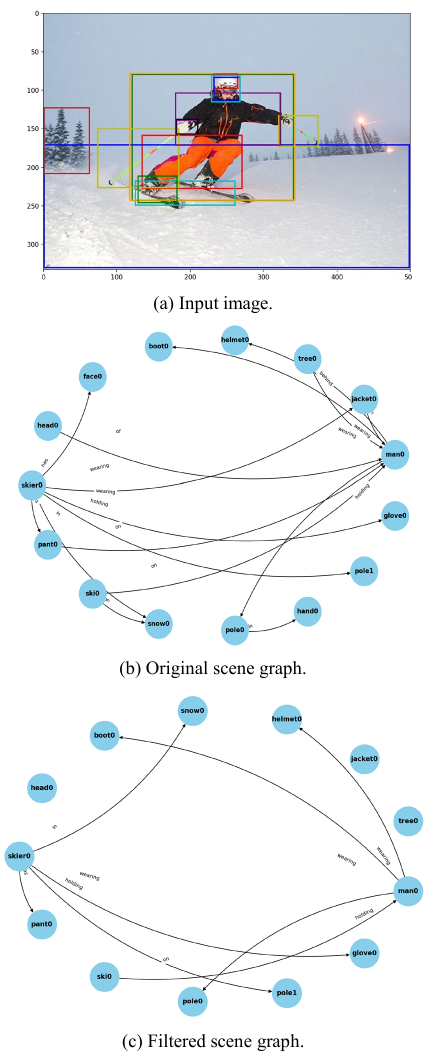}
\caption{Scene graph after two-stage PG-based compression.}
\label{fig:compressed_scenegraph_exp}
\end{figure}

In this simulation, we further present the workflow and effect of the proposed framework in real transmission scenarios. An example of extracted layout and full scene graph for a complex image from Visual Genome is shown in Fig.~\ref{fig:compressed_scenegraph_exp}(b). After applying the two-stage PG-based compression, the resulting significantly sparser scene graph is depicted in Fig.~\ref{fig:compressed_scenegraph_exp}(c). For this specific example, the ratio of omitted relation is approximately 55\%, and that of entity omission is about 7\%, demonstrating substantial reduction in semantic data.

\begin{figure}
    \centering
    \includegraphics[width=0.8\linewidth]{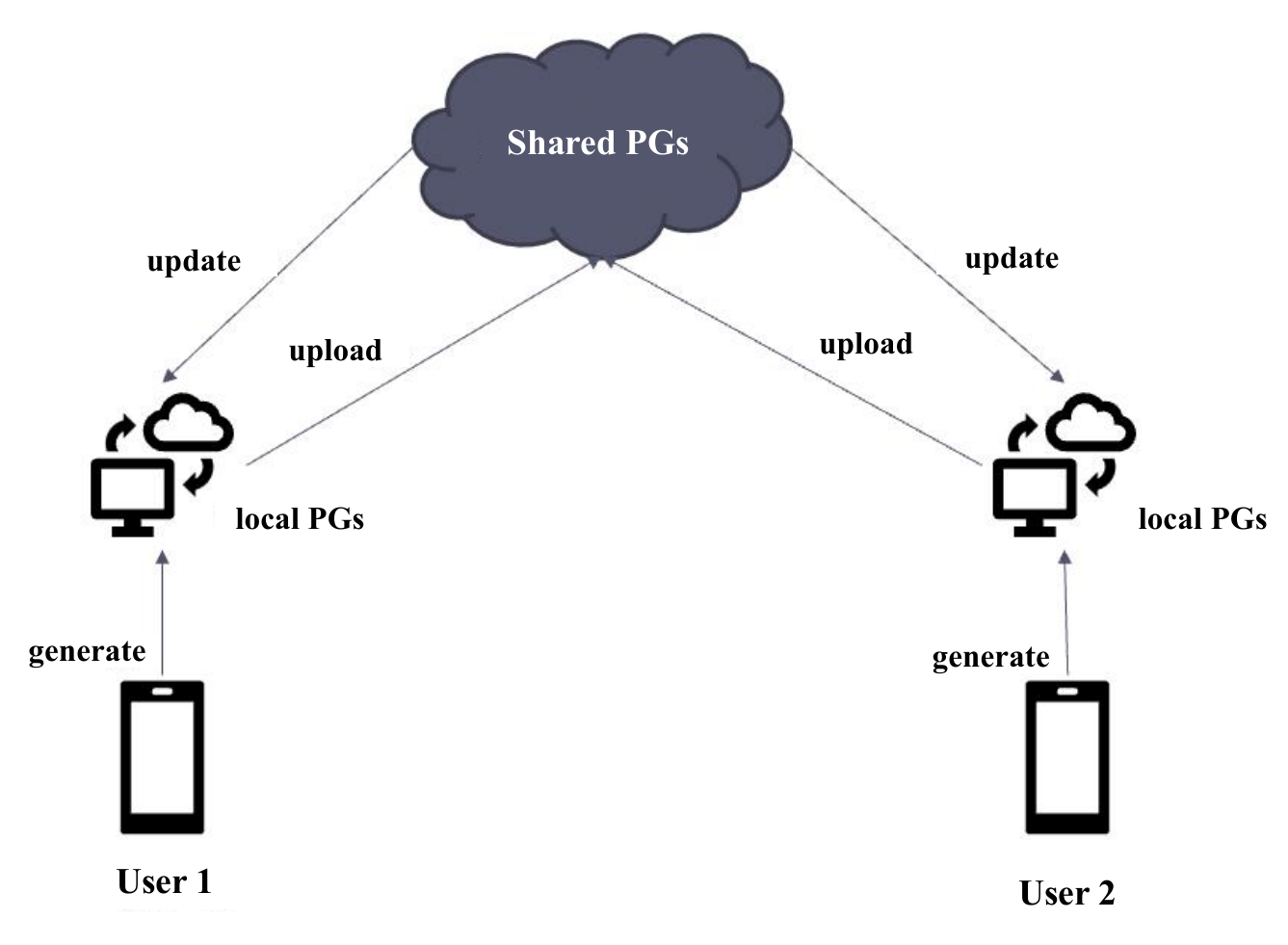}
    \caption{An illustration of the distributed transmission framework.}
    \label{fig:dtf}
\end{figure}

We propose a simple distributed communication framework as the simulation scenario. As illustrated in Fig.~\ref{fig:dtf}, two users (User1, User2) are involved in the communication process. They first generate the local PGs according to existing image data, and send the local PGs to a central server, where the sent data will be used to construct the shared PGs between two users, and download the shared PGs as the shared knowledge base during transmission. We assume that the image data of the two users has distinct data preferences (User1: 70\% human-related, 20\% animal, 10\% transport; User2: 70\% transport, 20\% animal, 10\% human). The local PGs were generated with varying numbers of local images  from 50 to 1000. A shared PG was created by combining their knowledge. 

\begin{figure}[t]
\centering
\includegraphics[width=\linewidth]{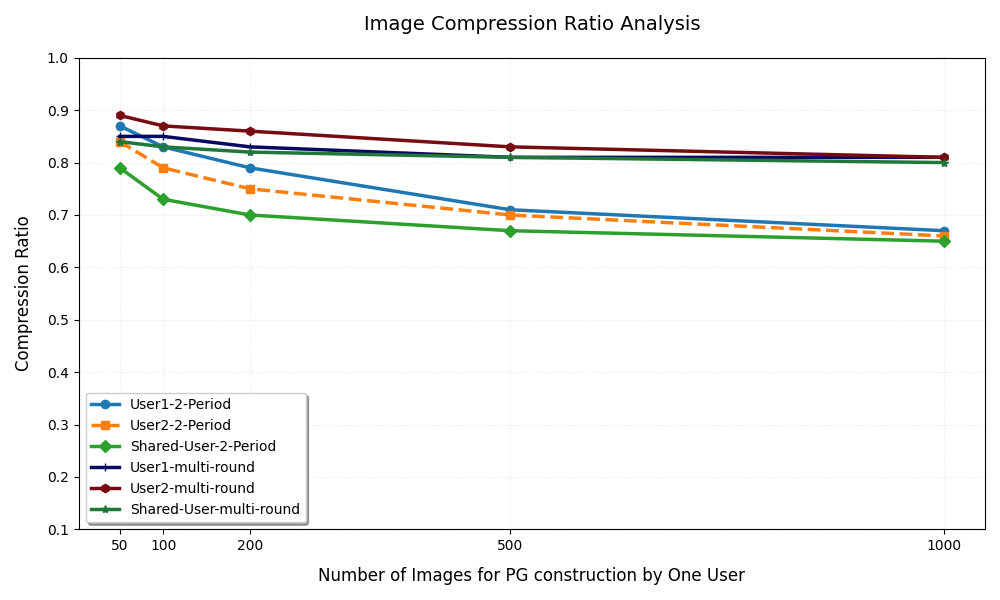}
\caption{Compression ratio versus number of training images per user for individual PGs (User1, User2) and the shared PG. Lower ratios indicate better compression.}
\label{fig:compression_ratio_comparison_exp}
\end{figure}

Fig.~\ref{fig:compression_ratio_comparison_exp} shows the average compression ratio achieved on a common test set of 100 images (40\% transport, 40\% human, 20\% animal).

In this simulation, we perform the 2-stage compression and the multi-round compression technique, respectively.
As shown in Fig.~\ref{fig:compression_ratio_comparison_exp}, increasing local training data improves the compression ratio (lower value is better) as the PG learns more data patterns. For all schemes, the shared PG consistently outperforms individual PGs, especially with smaller local datasets. This highlights the benefit of knowledge aggregation in the distributed setting, leading to more effective semantic compression. On the other hand, the compression rates achievable for multi-round compression are outperformed by that for 2-stage compression when user number is greater than 50. This may be result by the lack of entity filtering, as image contains rich entity information, while multi-compression focuses on relation filtering. 

\subsubsection{Transmission Throughput}
\begin{figure}[t]
\centering
\includegraphics[width=\linewidth]{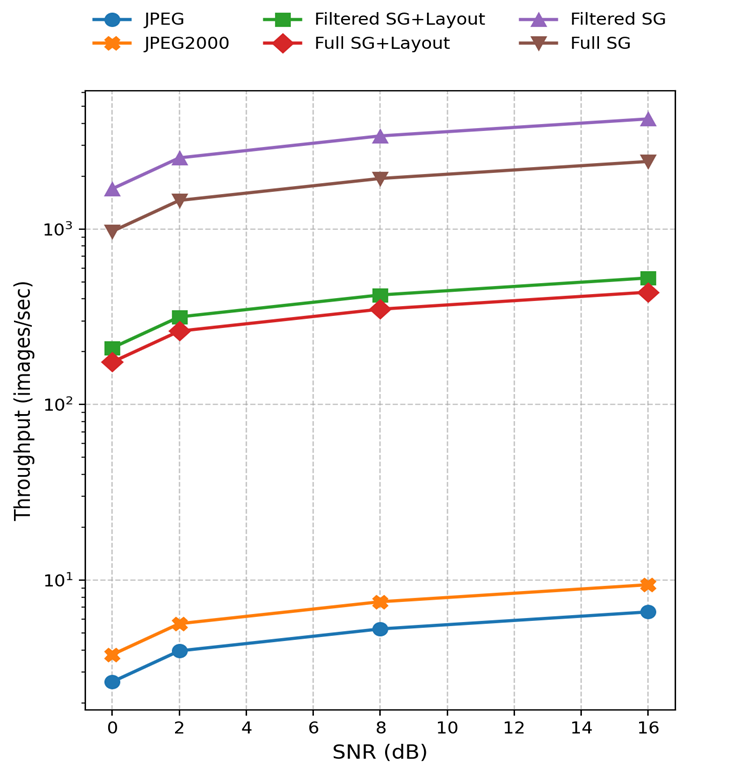}
\caption{Transmission throughput comparison for different data types across varying SNR values.}
\label{fig:throughput_comparison_exp}
\end{figure}

In this simulation, we compare the transmission throughput between traditional image transmission and the extracted or filtered scene graph (for simplicity, we only employ 2-stage filtering here). We simulated transmission over a 5G New Radio (NR) scenario with a 5MHz bandwidth, QPSK modulation, and LDPC channel coding (rates 1/3, 1/2, 2/3, 5/6 for SNRs 0, 2, 8, 16 dB respectively). Textual semantic information in encoded with UTF-8. To compare the performance of the proposed system, we perform 4 schemes in total:
\begin{itemize}
    \item Full SG: Image are processed into scene graphs. Only the scene graphs extracted will be used for transmission.
    \item Filtered SG: Image are processed into scene graphs. The scene graph will further go through relation filtering before transmission.
    \item JPEG: Images are compressed with JPEG algorithm before transmission.
    \item JPEG2000: Images are compressed with JPEG2000 algorithm before transmission. Comparing to JPEG, JPEG2000 is lossless concerning compression. As a result, the outcome image of JPEG2000 has higher resolution and larger data size.
    \item  Full SG+Layout/Filtered SG+Layout: These two schemes add the layout information to the corresponding base schemes (Full SG/Filtered SG) during transmission, which contains the relative position of known entities in the image.
\end{itemize}
Fig.~\ref{fig:throughput_comparison_exp} compares the throughput  for transmitting. Transmitting semantic information via scene graphs dramatically outperforms traditional image transmission. Scheme "Full SG" achieves over 200 times the throughput of scheme JPEG. PG-based compression scheme (`Filtered SG") further boosts this by approximately 1.64 times. Critically, "Filtered SG + Layout," which includes essential spatial information for reconstruction, enables transmission of hundreds of images per second even at 0dB SNR, whereas pixel-based methods struggle. This underscores the advantage of semantic transmission in bandwidth-limited, real-time applications.

\subsubsection{Image Reconstruction Quality}
\begin{figure}[t]
\centering
\includegraphics[width=\linewidth]{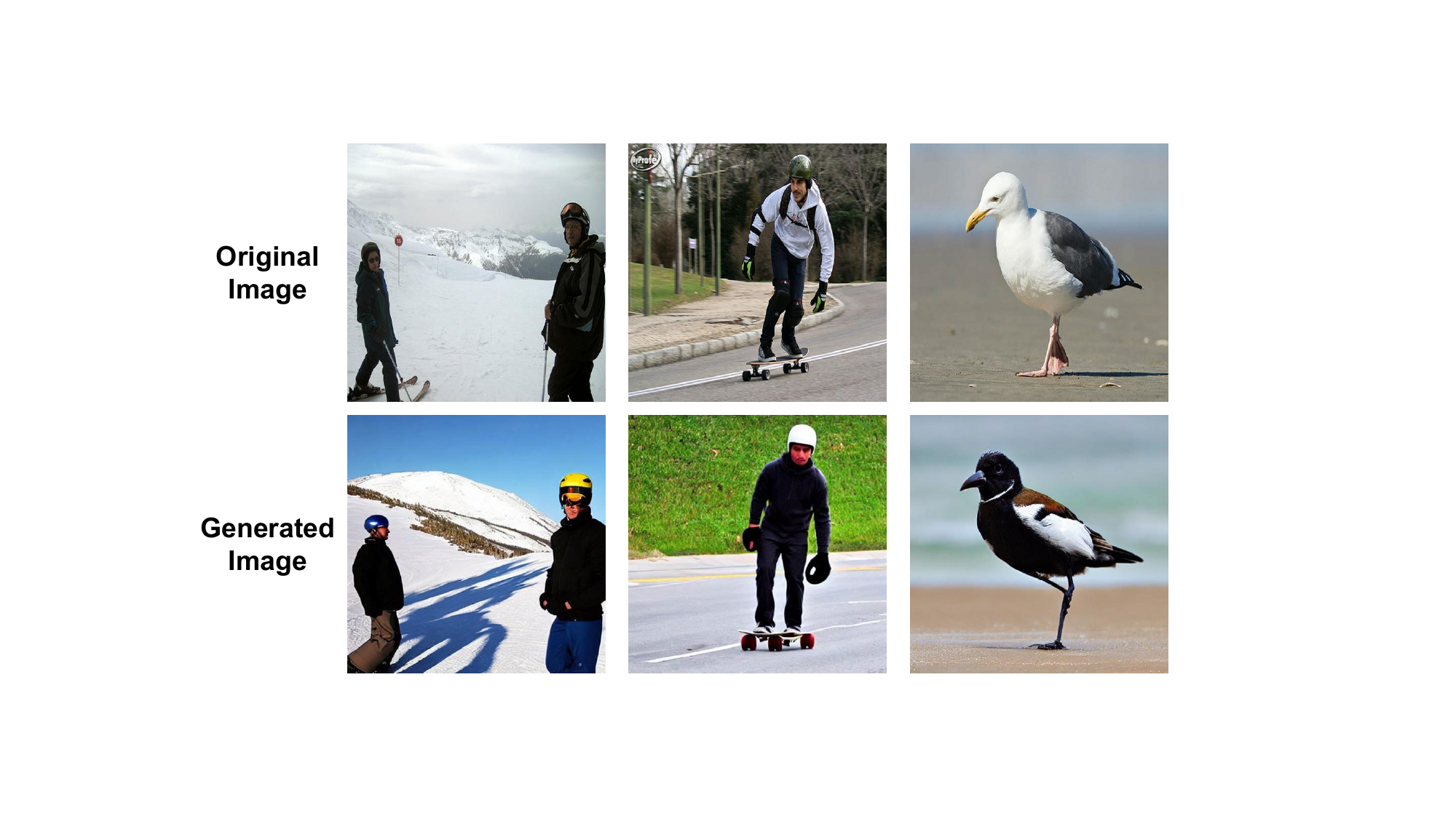}
\caption{Qualitative comparison of original images and images reconstructed by LDM from ``Full SG+Layout".}
\label{fig:image_comparison_exp}
\end{figure}

Images were reconstructed using an LDM conditioned on scheme `Full SG+Layout" and scheme `Filtered SG+Layout" (after PG-based recovery). Fig.~\ref{fig:image_comparison_exp} presents qualitative examples. While perceptible differences exist between original and generated images, the conceptual semantics are largely preserved.

\begin{figure}[t]
\centering
\includegraphics[width=\linewidth]{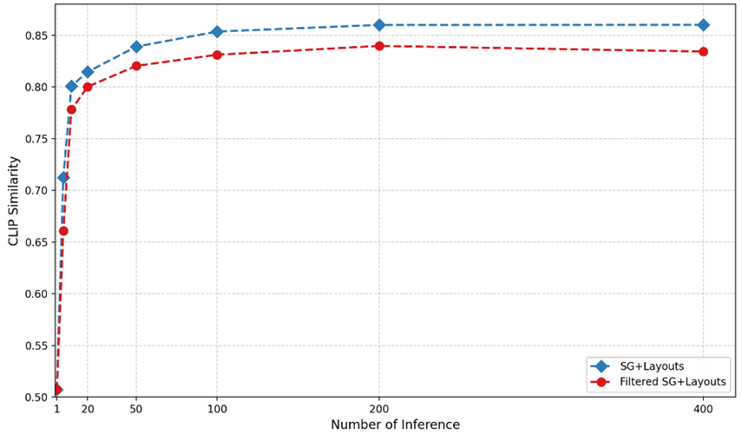}
\caption{CLIP similarity versus LDM inference steps for images.}
\label{fig:clip_similarity_exp}
\end{figure}

Quantitative evaluation using CLIP similarity (Fig.~\ref{fig:clip_similarity_exp}) shows that semantic consistency improves with more LDM inference steps, plateauing around 50 steps. Images reconstructed from "Full SG+Layout" achieve slightly higher CLIP scores than those from "Filtered SG+Layout," indicating that the PG-based recovery, while effective, might not perfectly restore all nuances, or that fuller semantic descriptions inherently lead to better alignment. Nevertheless, both achieve high CLIP scores ($>$0.8), demonstrating robust semantic preservation.

\begin{figure}[t]
\centering
\includegraphics[width=\linewidth]{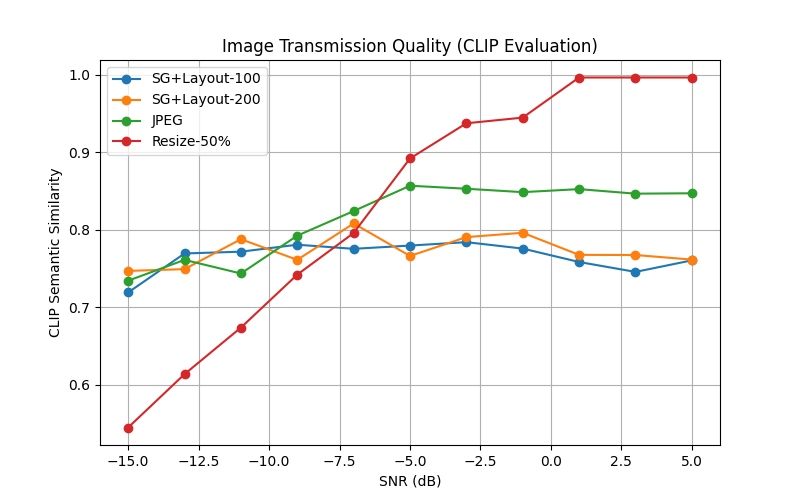}
\caption{SNR versus CLIP similarity.}
\label{fig:snr_v.s._clip}
\end{figure}

\begin{figure}[t]
\centering
\includegraphics[width=\linewidth]{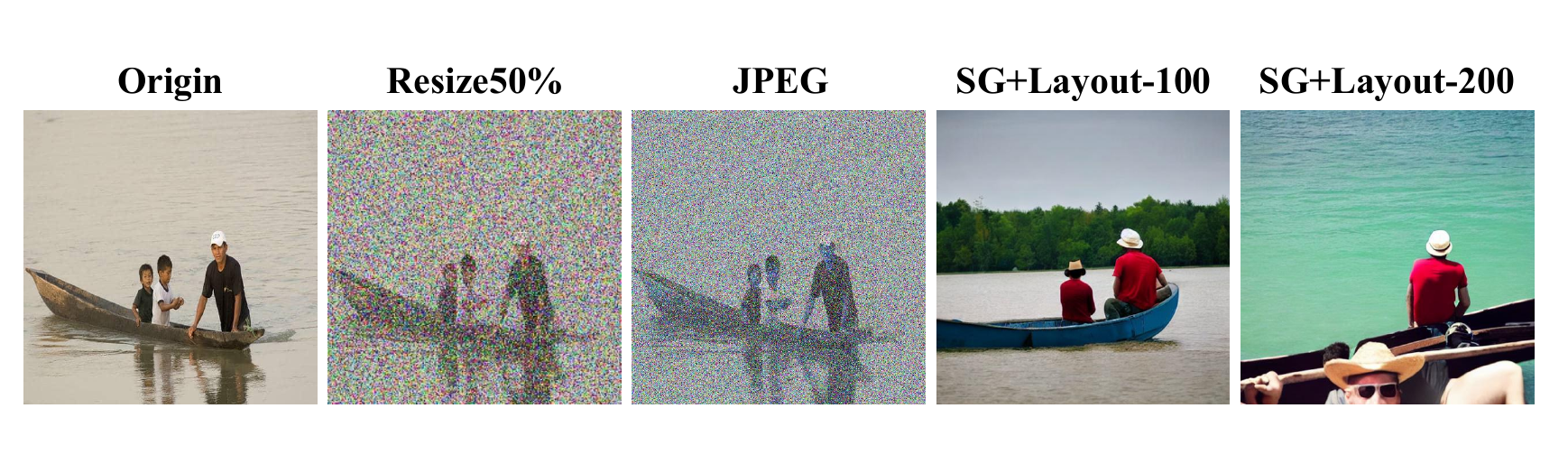}
\caption{Received / generated pictures when SNR=-15dB.}
\label{fig:snr_pic}
\end{figure}

In Fig.~\ref{fig:snr_v.s._clip}, we further compare the CLIP similarity of the outcome image after transmission (for scheme `JPEG' and scheme `Resize-50\%' it is the transmitted image, and for scheme `SG+Layout-100' and scheme `SG+Layout-200' it is the generated image) and the original one. In order to examine the effect of probabilistic semantic communication without complex anti-noise techniques, we conduct the experiment using simple BPSK (binary phase shift keying) . The carrier wave frequency is set to 5000 and the transmitting speed is fixed at 50 bits per second. For schemes transmitting scene graphs and layout, the scene graph adopted is filtered by the proposed 2-stage method and we set distinct inference steps for LDM (100 or 200) to compare the outcome CLIP similarity. As for scheme `Resize-50\%', it resizes the image into half the size it is before transmission. The experiment is conducted on single image transmission. As can be shown in Fig.~\ref{fig:snr_v.s._clip}, the outcome CLIP similarities of SG-implemented schemes are closed to or even surpass normal compression methods in low-SNR environments. This is due to the fact that the size of the transmitted data is minimal, which results in a reduced probability of noise pollution. This advantage can be shown in Fig.~\ref{fig:snr_pic}, which demonstrates the received/generated pictures of all schemes when SNR=-15dB. After regeneration, the generated image has superior definition, and still captures the key semantic components (man on a boat / boat in a river) in the origin image. In high-SNR environments, the CLIP similarity of scheme `Resize-50\%' is equal to 1, since the transmitted image is almost identical to the original one. However, this level of accuracy is accompanied by a substantial increase in data size. It is notable that in high-SNR environment, the generated image's CLIP similarity lags behind, but the transmit throughout is still prior than the other schemes, as shown in Fig.~\ref{fig:throughput_comparison_exp}.

Table \ref{tab:lpips_comparison_exp} shows the LPIPS scores for filtered and not filtered schemes measured at 50 inference steps. When using LPIPS metric, lower score indicates more similarity. The result favors `Full SG+Layout", suggesting better perceptual similarity due to more detailed input semantics. The relatively high LPIPS values overall indicate that while semantic concepts are captured, fine-grained visual details present in the original pixel domain may be lost or altered when reconstructing solely from high-level scene graphs and layouts.

\begin{table}[htbp]
\caption{LPIPS Metric Comparison at 50 Inference Steps}
\label{tab:lpips_comparison_exp}
\centering
\begin{tabular}{|l|c|c|}
\hline
Input for LDM & LPIPS (AlexNet) & LPIPS (VGG16) \\
\hline
Filtered SG + Layouts & 0.619 & 0.637 \\
Full SG + Layouts & 0.573 & 0.609 \\
\hline
\end{tabular}
\end{table}

In summary, the experiments validate the proposed PG-based semantic communication system. It offers substantial compression (average 64.7\% reduction in scene graph size for the test set using the shared PG with 1000 images per user), leading to significant throughput gains, especially in low SNR conditions. The PG-based recovery mechanism effectively reconstructs semantic information, enabling high-quality image generation with strong semantic consistency, although with some loss in pixel-level perceptual fidelity compared to original images.

\subsection{Multi-Round Semantic Compression Simulation}
To examine the power consumption theoretical assumption in this paper, we employ the compression algorithm on a test dataset and view the impact of multi-round semantic compression.

The dataset is collected from \cite{sensordataset}. The data is generated from a test house, where 2 residents lived inside. There were motion sensors placed in the house, denoted as `M01', `M02',...`M49', and 3 other sensors `AD1-A', `AD1-B' and `AD1-C'. 

\begin{figure}
    \centering
    \includegraphics[width=\linewidth]{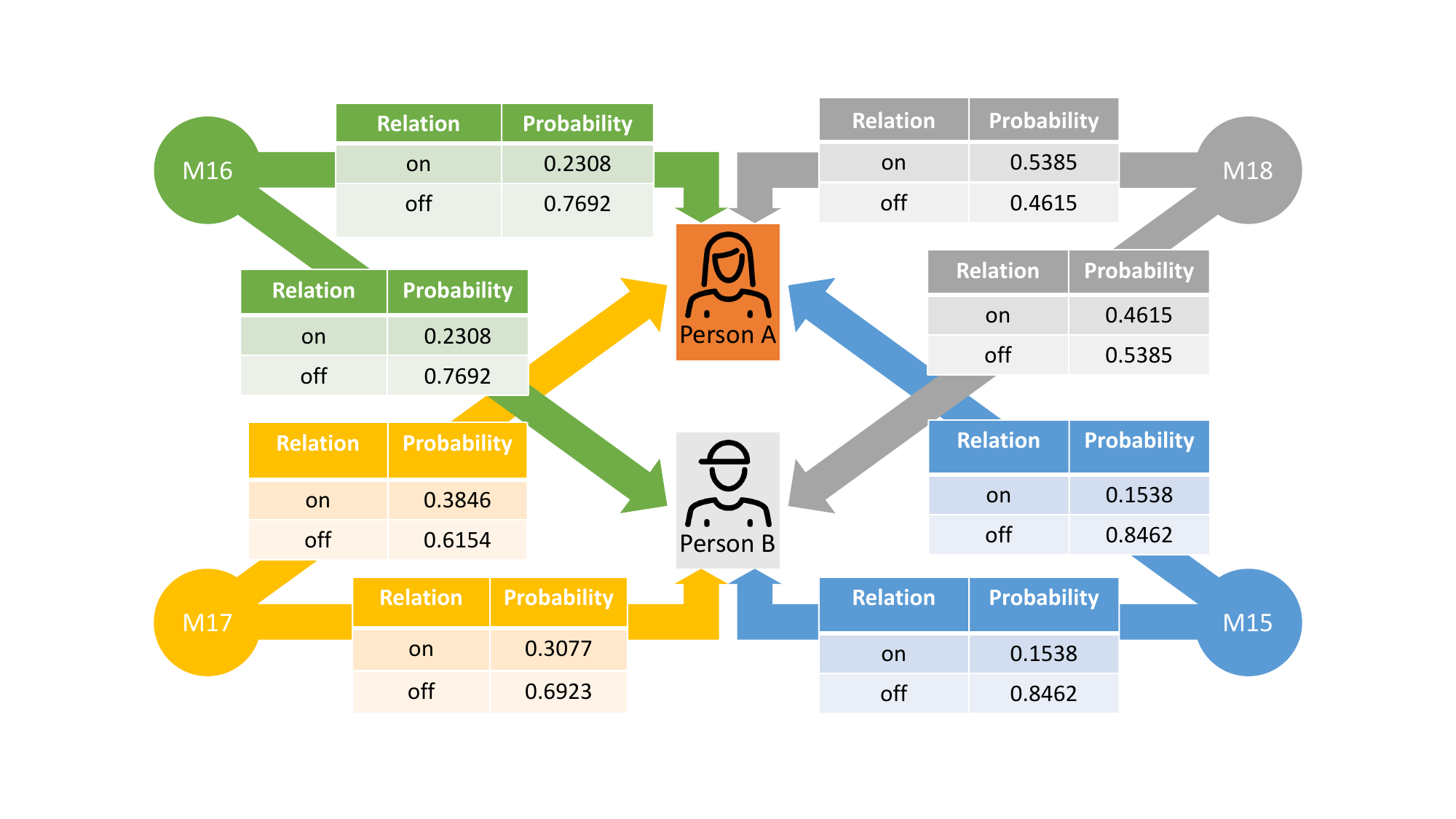}
    \caption{An illustrative figure of partial extracted probability graph.}
    \label{fig:pg}
\end{figure}

We define the relationship between sensors and a person as either `on' (triggered) or `off' (not triggered). Data is sampled based on specific time ranges, with each sample representing a one-minute interval (e.g., all records occurring at 19:20 on 6/30). If a sensor is triggered by a person, the relationship between that sensor and the person is marked as `on' for that sample, while the relationships between all other sensors and the same person are marked as `off'. This relationship remains consistent until new records are processed, where any sensors previously marked as 'off' for that person may be triggered, updating their status to `on' for the current sample. Within a single sample, the relationship between a specific sensor and a person is unique—either `on' or `off'.An illustrative figure of partial extracted probability graph is given in Fig.~\ref{fig:pg}.

\begin{figure}
    \centering
    \includegraphics[width=\linewidth]{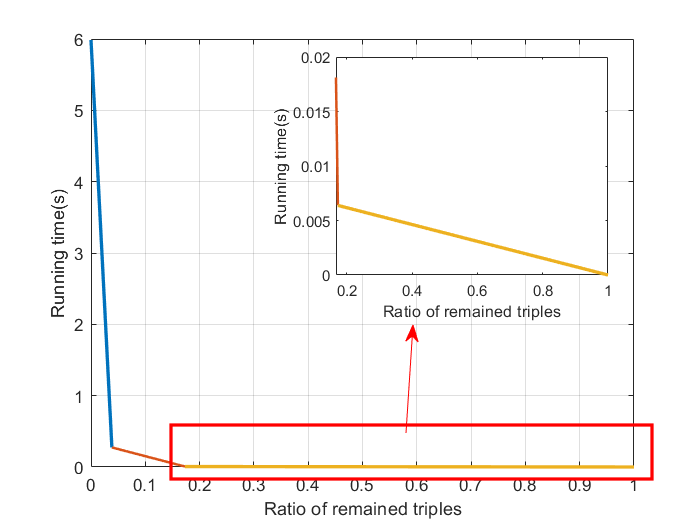}
    \caption{Running time versus ratio of remained triples.}
    \label{fig:msc}
\end{figure}

As shown in Fig.~\ref{fig:msc}, the power consumption nature is closed to our theoretical assumption. It must be noted that the test data has significant hidden relativity, and presents a typical distribution according to the shared probability graph. For practical use, the minimum compression ratio depends on the distribution of both the transmit data and the shared probability graph.
\section{Conclusion}
\label{sec:conclusion}
This paper introduced a distributed image semantic communication system that leverages probabilistic graphical models as a shared knowledge base to enhance transmission efficiency. We focused on the representation of high-level image semantics using scene graphs and developed PG-based algorithms for their compression and recovery. The core idea is that statistical regularities in semantic structures, learned by the PG from a large dataset, can be exploited to eliminate predictable information at the transmitter and reliably infer it at the receiver. A distributed PG construction approach was also proposed to allow for collaborative knowledge sharing and improved adaptability across diverse user contexts. For further deployment, we also demonstrated a multi-round semantic compression approach and analyzed its computation load. The practicality and power consumption nature were investigated through simulation.

Experimental results demonstrated the efficacy of our approach. The PG-based compression significantly reduced the data volume of scene graphs, leading to orders-of-magnitude improvements in transmission throughput compared to traditional image compression techniques, particularly under bandwidth-constrained scenarios. The reconstructed images, generated using a Latent Diffusion Model conditioned on the recovered semantics, exhibited high semantic consistency with the originals, as measured by CLIP similarity. While some loss in fine-grained perceptual detail was observed (via LPIPS scores), the ability to convey core semantic content with drastically reduced bandwidth is a key advantage. The distributed PG training showed benefits, with shared knowledge leading to better compression than individual, smaller PGs.

\bibliographystyle{IEEEtran}
\bibliography{ref}

\end{document}